\documentclass[12pt,3p]{elsarticle}
\usepackage{graphicx}
\usepackage{amssymb}
\usepackage{lineno}
\usepackage{bm}

\journal{Chemical Physics}

\newcommand{\rmi}{{\rm i}}
\newcommand{\rme}{{\rm e}}
\newcommand{\rmd}{{\rm d}}

\def\bra#1{\mbox{$\langle#1|$}}
\def\ket#1{\mbox{$|#1\rangle$}}

\makeatletter
\long\def\@ympar#1{%
  \@savemarbox\@marbox{\small #1}%
  \global\setbox\@currbox\copy\@marbox
  \@xympar}
\makeatother

\begin{document}

\begin{frontmatter}

\title{Energy flow in the Photosystem~I supercomplex: 
comparison of approximative theories with DM-HEOM}

\author[1,2]{Tobias Kramer}
\ead{kramer@zib.de}
\author[1]{Matthias Noack}
\author[3,4]{Jeffrey R Reimers}
\author[1]{Alexander Reinefeld}
\author[1]{Mirta Rodr\'iguez}
\author[5]{Shiwei Yin} 
\address[1]{Zuse Institute Berlin (ZIB), Takustr.\ 7, 14195 Berlin, Germany}
\address[2]{Department of Physics, Harvard University, 17 Oxford Street, 02138 Cambridge, Massachusetts, United States}
\address[3]{School of Mathematical and Physical Sciences, University of Technology Sydney, Australia}
\address[4]{International Centre for Quantum and Molecular Structure, School of Physics, Shanghai University, Shanghai 200444, People's Republic of China}
\address[5]{Key Laboratory for Macromolecular Science of Shaanxi Province, School of Chemistry \& Chemical Engineering, Shaanxi Normal University, Xi'an City, 710062, People's Republic of China}

\begin{abstract}
We analyze the exciton dynamics in Photosystem~I from {\it Thermosynechococcus elongatus} using the distributed memory implementation of the hierarchical equation of motion (DM-HEOM) for the 96 Chlorophylls in the monomeric unit.
The exciton-system parameters are taken from a first principles calculation.
A comparison of the exact results with F\"orster rates and Markovian approximations allows one to validate the exciton transfer times within the complex and to identify
deviations from approximative theories.
We show the optical absorption, linear, and circular dichroism spectra obtained with DM-HEOM and compare them to experimental results.
\end{abstract}


\end{frontmatter}

The energy transfer dynamics in photosynthetic light harvesting complexes (LHC) is understood as a transfer of electronic excitation from the antenna pigments which absorb visible light at high energy into the reaction center (RC) where the chemical reactions take place \cite{Reimers2016}. 

The electronic excitation of individual pigments delocalizes into exciton states due to dipole-dipole interactions. 
The coupling of the exciton to the molecular motion and vibration leads to thermal dissipation, which eventually directs the exciton to the lower energy states and the RC.

The key for understanding the transport processes in pigment networks of realistic system sizes and transferring results from natural to artificial systems is to determine the energy pathways and functional roles of the different subsystems \cite{Sobolewski2008}. 

Here, we track the energy flow in the different pathways across the Photosystem~I (PS~I) complex and analyse the role of dissipation in the energy transfer dynamics.
The transfer time through the PS~I to the reaction center depends critically on the strength of the inter-pigment dipole-dipole-dominated couplings, in addition to the vibrational thermal dissipation.  
Approximative theories of energy transfer for larger complexes carry along unknown errors, since the couplings between the pigments are spanning a wide parameter range, giving rise to both, localized and delocalized states.

Therefore, we base all computations on the numerically exact hierarchical equations of motion (HEOM) method \cite{Tanimura1989,Ishizaki2009,Kreisbeck2011}, which captures the system-environment dissipation and decoherence, while retaining coherences and non-Markovian effects in a non-perturbative manner.
For computing the dynamics and spectra of PS~I an efficient parallelization of the HEOM method is required, which is provided by the Distributed Memory DM-HEOM software package \cite{Noack2018,Kramer2018a}.
 
The article is organized as follows:
In Sec.~\ref{sec:Model} we introduce the Frenkel exciton model of the PS~I complex and we briefly review in Sec.~\ref{sec:Dynamics} the HEOM method and the approximative rate equations.
Sec.~\ref{sec:Pop} analyzes the population dynamics in the PS~I complex for the different methods.
We identify the various pathways within the complex and provide a comparison to previous results in the literature based on approximative methods \cite{Byrdin2002,Sener2002,Yang2003,Bruggemann2004a,Abramavicius2009a}. 
Sec.~\ref{sec:Spectra} is devoted to the description of the optical absorption and circular dichroism spectra.
The article ends with some concluding remarks in Sec.~\ref{sec:Discussion}.

\section{Photosystem I}\label{sec:Model}

The structure of Photosystem~I from the thermophilic cyanobacterium {\it Thermosynechococcus elongatus} has been resolved by X-ray crystallography to 2.5~{\AA} resolution \cite{Jordan2001}. 
The PS~I appears in a trimeric structure, each monomeric unit consisting of 12 protein subunits, with 96 cholorophyll (Chl) molecules, 22 carotenoids, three iron-sulfur clusters, two phylloquinones, four lipids, $201$ water molecules, and at least one metal ion \cite{Canfield2006}.

\begin{figure}[h]
\centering\includegraphics[width=0.6\textwidth]{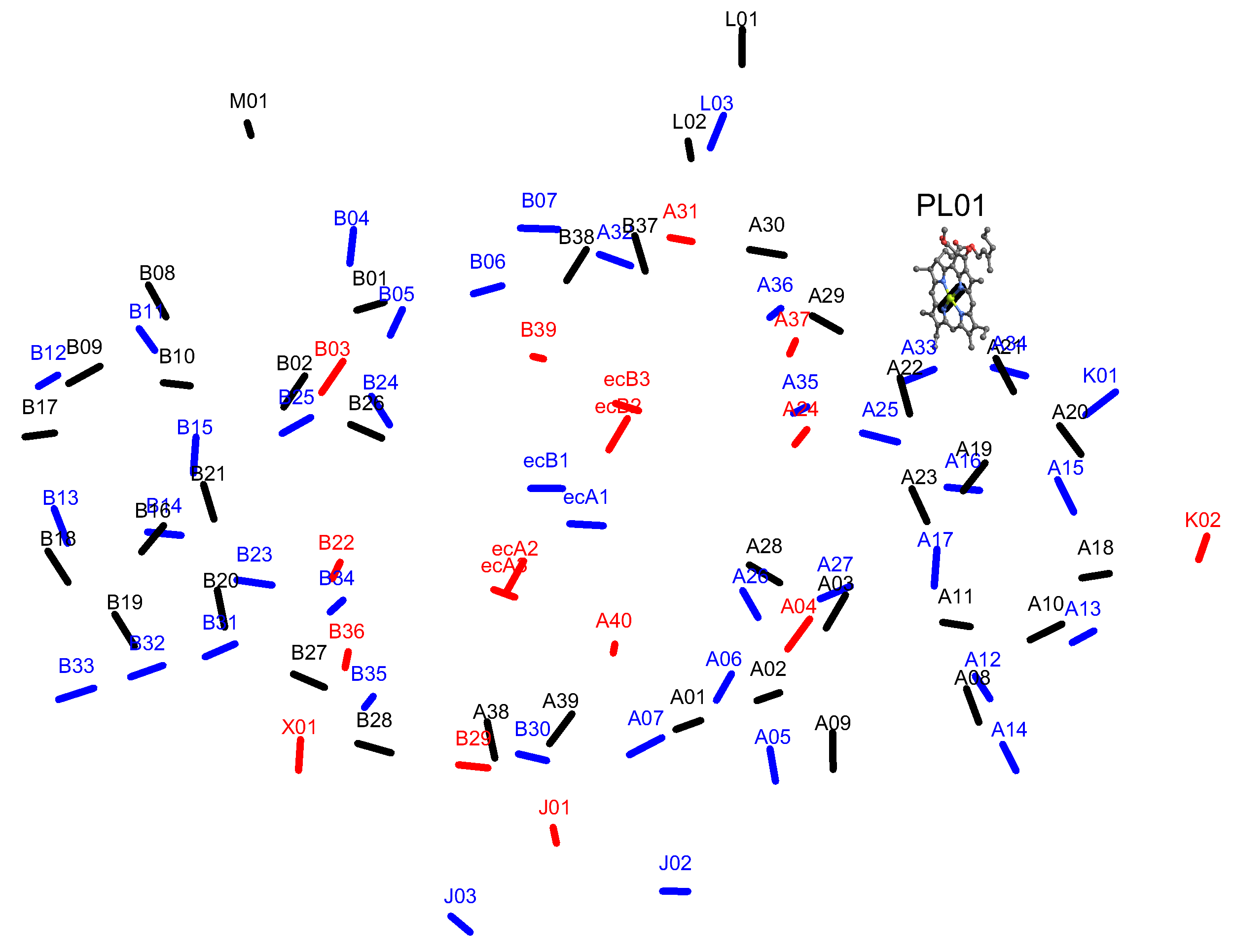}
\caption{Arrangement of the $96$ PS~I Chl pigments.
The lines connect the nitrogen $N_B$ $N_D$ positions, which are taken as the direction of the transition dipole moment.
(Optimized structure 1JB0 from \cite{Jordan2001,Yin2007}).
External pigment PL01 (highlighted) is taken as initial condition in the dynamics calculations.
}\label{fig:psi}
\end{figure}

Knowledge of the detailed arrangement of the Chls is a big step forward towards a more realistic description of the excitation energy transfer in PS~I \cite{Fromme2001}.
The antenna of PS~I consists of 90 Chl pigments surrounding a central six Chl core, which was identified as the reaction center where charge separation takes place. 
The top view of the monomer PS~I structure (Fig.~\ref{fig:psi}) shows that the two largest subunits of PS~I (PsaA and PsaB) bind most of the antenna pigments (chlorophylls A1-A40 and B1-B39 respectively) in two interconnected stromal and lumenal ring structures slightly twisted against each other.
The central RC Chls are organised in A-B pairs, including the primary or special electron pair P700 (chlorophylls ecA1, ecB1), with two of the pairs located in a middle layer between the stromal and lumenal antenna rings \cite{Jordan2001,Byrdin2002}. 
In addition, 11 antenna Chls are hosted by the outer subunits. 
An outer to inside transport is proposed for the excitation energy transfer within the PS~I monomer based on previous calculations \cite{Byrdin2002,Sener2002,Damjanovic2002,Yang2003,Bruggemann2004a,Yin2007,Abramavicius2009a}.

\subsection{Frenkel exciton model of PS~I}\label{sec:ReimersModel}

The Frenkel exciton \cite{May2004} system is written as
\begin{equation}
\label{eq:h-full}
H=H_{\rm g}+H_{\rm ex}+H_{\rm bath}+H_{\rm ex-bath},
\end{equation}
where $H_{\rm g}=\varepsilon_0\ket{0}\bra{0}$ represents the ground state Hamiltonian (ground state energy $\varepsilon_0$), $H_{\rm ex}$ denotes the excitation energies and interactions of the pigments, and $ H_{\rm bath}$ models the effect of the surrounding protein environment as a vibrational bath coupled to each pigment by $H_{\rm ex-bath}$.  
The excitonic Hamiltonian $H_{\rm ex}^{\rm site}$ for a system of $N$ constituents (``sites'') is parametrized as
\begin{equation}
\label{eq:h-ex} 	
H_{\rm 0}^{\rm site} =\sum_{m=1}^{N}\varepsilon_m^0\ket{m}\bra{m}+\sum_{n\neq m}J_{mn}\ket{m}\bra{n},\quad
H_{\rm ex}^{\rm site} =H_{\rm 0}^{\rm site}+\sum_{m=1}^{N}\lambda_m\ket{m}\bra{m},
\end{equation}
where we introduce the energy $\varepsilon_{m}=\varepsilon_{m}^{0}+\lambda_m$, which consists of the zero phonon energy  $\varepsilon_{m}^{0}$ shifted by the reorganization energy $\lambda_m$, and the coupling matrix elements $J_{mn}$. 

The molecular data $\varepsilon_m^0$ and $J_{mn}$ are evaluated \cite{Yin2007} using an a priori computational approach.
This is based on a 150000-atom structure for the PS~I trimer optimized from the original X-ray coordinates \cite{Jordan2001} using the PW91 density functional \cite{Perdew1992}, the 6-31(+)G* basis set \cite{Hehre1972}, and using a linear-scaling technique \cite{Canfield2006}.
This optimization produced large changes to the coordinates of the critical chlorophyll special pair, plus required optimization of all chlorophylls, as well as many other changes not related to the exciton model.
At this geometry, $\varepsilon_m^0$ and $J_{mn}$ are evaluated using CAM-B3LYP \cite{Kobayashi2006} with the 6-31G* basis \cite{Hehre1972}.
These calculations evaluated the energies and transition moments of all pairs of chlorophyll molecules up to $20$~\AA~
in separation considered in isolation, interpreting them in terms of a basic two-level exciton model.
This first-principles approach includes all Forster, Dexter, and higher-order contributions.
CAM-B3LYP was selected as it was the first density-functional method shown to be able to deal with charge-transfer effects critical to chlorophyll spectroscopy \cite{Cai2006}.
It predicts details of the absorption and emission spectra of chlorophylls with great accuracy \cite{Ratsep2011} and predicted all key features of what later was established as the assignment of the Q-band spectra of the chlorophylls \cite{Reimers2013}.
This method is therefore expected to yield a qualitatively realistic description of the zero-phonon energies and the coupling matrix elements, and indeed the results obtained are broadly consistent \cite{Yin2007} with empirical deductions \cite{Byrdin2002,Sener2002,Yang2003,Bruggemann2004a,Abramavicius2009a,Damjanovic2002}.
Other approaches include the Poisson-TrEsp method and Charge Density Coupling \cite{Adolphs2010} applied to PS~I.
Observed data can be very sensitive to small changes in these parameters, however, and this a priori calculated data may require subtle improvements.  One simplification of the primary data is utilized in the current application: the PS I trimer presents a cyclic network of 286 chlorophylls, but herein only a single monomer containing 96 chlorophylls is explicitly represented, with periodic boundary conditions applied to connect one side of this monomer back to its other side.

The vibrational environment consists of $N$ uncorrelated baths $H_{{\rm bath},m}=\sum\limits_{i}\hbar\omega_{m,i}(b_{m,i}^{\dagger}b_{m,i}+\frac{1}{2})$ of harmonic oscillators of frequencies $\omega_{m,i}$, with bosonic creation and annihilation operators $b_{m,i}$. 
The oscillator displacement of each bath $(b^{\dagger}_{m,i}+b_{m,i})$ is coupled to the exciton system by
\begin{equation}
H_{\rm ex-bath}=\sum_m \ket{m}\bra{m} \otimes \sum_i \hbar\omega_{m,i} d_{mi}(b^{\dagger}_{m,i}+b_{m,i}),
\end{equation}
where $d_{mi}$ denotes the coupling strength related to the spectral density $J_m(\omega)=\pi\sum_i \hbar^2\omega_{mi}^2d_{mi}^2\delta(\omega-\omega_i)$.\\
The spectral density is connected to the reorganization energy 
\begin{equation}
\lambda_m=\int_0^\infty \frac{J_m(\omega)}{\pi\omega}\rmd\omega.
\end{equation}
Here, we consider a Drude-Lorentz spectral density
\begin{equation}
\label{eq:spectral_density_DL}
J_{m}(\omega)=2\frac{\lambda_{m}\omega\nu_m}{\omega^2+\nu_m^2}\;,
\end{equation}
with bath correlation time $\nu_{m}^{-1}$.
In principle, the vibrational structure of the $Q_y$ band should also be included in the model, as also should be exciton transfer including the chlorophyll $Q_x$ bands, processes for which detailed data is now available \cite{Canfield2006}.
However, these processes occur at energies  $800$-$2000$~cm$^{-1}$ above the $Q_y$ origin transitions considered herein, out of the energy range of primary interest. 
Test calculations performed including these effect indicate that indeed they may be safely neglected.
Some authors assign to a group of ``red'' chlorophylls a larger reorganization energy \cite{Yang2003} based on hole-burning data \cite{Gillie1989,Zazubovich2002} while others do not treat the spectral density of these chlorophylls differently \cite{Adolphs2010}.
For simplicity and since to our knowledge no computed reorganization energies are published, we do not consider variations in reorganization energies between the pigments.

\section{Theories for exciton dynamics}\label{sec:Dynamics}

The exciton dynamics of the molecular complex is described by the reduced density matrix of the system.
For large complexes (PS~I with 96 pigments) and for transfer times exceeding tens of picoseconds, the time-propagation of the density matrix is computationally demanding.
Previous simulations are based on approximative methods, often reducing the density matrix to a population vector \cite{Byrdin2002,Yang2003,Renger2006}.
In the following we study the deviation of the population rate equations from the exact HEOM result.
The spectral density of the molecular vibrations, Eq.~(\ref{eq:spectral_density_DL}), is taken to be the same for all Chls with $\lambda=35$~cm$^{-1}$, $\nu^{-1}=50$~fs.

\subsection{Rate equations, F\"orster theory}

For PS~I, a generalized F\"orster theory combined with modified Redfield dynamics for weakly coupled parts of the system has been applied \cite{Yang2003a,Renger2006}.
The modified Redfield method is not directly comparable to HEOM, since it only computes the populations in energy representation, but neglects any coherences.
For a direct comparison with the HEOM reference calculation, we use the F\"orster expression for the rates $\mathbf{R}$ in site basis
\begin{equation}
R_{m,n}=2|J_{mn}|\Re \int_0^\infty \rmd t F_m^*(t) A_n(t),
\end{equation}
with 
\begin{eqnarray}
A_n  (t)&=&\exp[-\rmi (\epsilon_n^0+\lambda_n) t-g_n(t)],\\
F_m^*(t)&=&\exp[+\rmi (\epsilon_m^0-\lambda_m) t-g_m(t)],\\
g_m(t)&=&-\frac{1}{2\pi}\int_{-\infty}^\infty\rmd\omega \; \frac{J_m(\omega)}{\omega^2}
\left(1+\coth(\beta\hbar\omega/2)\right)\left(\rme^{-\rmi\omega t}+\rmi\omega t-1\right).
\end{eqnarray}
For the Drude-Lorentz spectral density considered here, the analytic expression of $g(t)$ reads in terms of Lerch's transcendent $\Phi(z,s,a)$ and the Harmonic number function $H(z)=\psi_{\rm digamma}(z+1)+\gamma_{\rm Euler-Mascheroni}$
\begin{eqnarray}\label{eq:gexact}
g(t)&=&
\frac{\lambda  \rme^{-\frac{(b+1) \nu  t}{b}}}{\pi  \nu }
\bigg(
\rmi ( \rme^{\frac{(b+1) \nu  t}{b}} (H(-b) (\nu  t-1)-H(b) (\nu  t+1)))\nonumber\\
&&
-\rmi \rme^{\nu  t} (\Phi (\rme^{-\frac{t \nu
   }{b}},1,1-b)+\Phi (\rme^{-\frac{t \nu }{b}},1,b+1)+2 \rme^{\frac{\nu  t}{b}} \log (1-\rme^{-\frac{\nu 
   t}{b}}))
\nonumber\\
&&
+\frac{2 \pi  \left(\rme^{\nu  t} (\nu  t-1)+1\right) \rme^{\frac{\nu  t}{b}}}{-1+\rme^{2 \rmi \pi  b}}
\bigg)\\
b&=&\frac{\nu\hbar}{2 \pi k_B T}.
\end{eqnarray}
The population dynamics in F\"orster theory is then given by
\begin{eqnarray}
\rho_{mm}(t)&=&\rho_{mm}(0) \rme^{\mathbf{K} t},\\
K_{\alpha\alpha}&=&-\sum_{\gamma=1,\gamma\ne\alpha}^{N} R_{\gamma\alpha},\\
K_{\alpha\beta}&=&R_{\alpha\beta}, \quad (\alpha\ne\beta)
\end{eqnarray}

\subsection{Hierarchical equations of motion (HEOM)}

The HEOM method \cite{Tanimura1989} has been used as reference method for studying energy transfer processes, since it accurately covers a wide parameter range of couplings and temperatures \cite{Kramer2018a}.
HEOM is expressed as a system of coupled differential equations for the time evolution of $N_{\rm matrices}$ auxiliary density matrices $\sigma_u$ of dimensions $N\times N$ (for PSI~I: $N=96$).
The matrices are arranged in layers of increasing depth $D$, connected by vertices with "+" upward and "-" downward links between the matrices in each layer.
\begin{equation}
\frac{\rm d \sigma_u}{\rmd t}
=-\frac{\rmi}{\hbar}\left[H, \sigma_u\right]
+\sum_{\rm baths} A \sigma_{u}
+\sum_{\rm baths} B \sigma_{u_{+}}
+\sum_{\rm baths} C \sigma_{u_{-}}.
\end{equation}
Explicit expressions for the operators $A,B,C$ are given in \cite{Tanimura2006} and \cite{Kramer2018a}, Eqs.~(12-36).
The layer size increases with the layer number, while the top-layer contains only a single density matrix $\sigma_0$, which represents the desired reduced density matrix.
The layer structure is generated by the Taylor expansion of the exponentially decaying bath correlation function and denotes increasingly higher derivatives.
For typical light harvesting complexes, convergence is reached upon inclusion of the first few ($D=2$-$3$) layers at $T=300$~K \cite{Kramer2018a}.
The number of matrices required for solving the HEOM system at layer depth $D$ increases with the system size $N$, the number of vibrational baths $B$, and the number of Matsubara modes $M$ required to describe lower temperatures.
It is given by the binomial
\begin{equation}
\label{eq:Nmatrices}
N_{\rm matrices}=
\left(\begin{array}{c}
M B+D\\
M B
\end{array}\right).
\end{equation}
For the PS~I complex, this results for $B=96$, $M=1$, $D=3$ in 156\,849 matrices, for $D=4$ in 3\,921\,225 matrices.
We use the Distributed Memory (DM-HEOM) software package \cite{Noack2018,Kramer2018a} to compute the PS~I complex, which allows to distribute the computation and memory across multiple compute nodes.
Previous single node implementations of HEOM \cite{Kreisbeck2011,Strumpfer2012a} were limited in the treatable system size by the high memory requirements to store millions of matrices and by the computation time required by the matrix-matrix multiplications.
One of the largest molecular complexes considered with the QMaster implementation of HEOM \cite{Kreisbeck2014} has been a model of the PS~II supercomplex \cite{Kreisbeck2015} comprising 93~pigments.

DM-HEOM overcomes these limitations by distributing the data structures and computations, thus providing scalability from a single computer to hundreds of compute nodes of a supercomputer.
This enables the simulation of systems with more than 100 sites and 3 or more hierarchy layers. 
The PS~I hierarchy matrices have a memory footprint of 129~GiB of memory for $D=3$ layers in conjunction with a fourth-order Runge-Kutta solver.
For $D=4$ hierarchy layers, the memory demand increases to 3231~GiB, reflecting the growth of the hierarchy with each added layer.
Simulating $1000 \times 1$~fs steps of the population dynamics of the PS~I complex with 3 layers takes about 22 minutes on 256 compute nodes of a Cray XC40 supercomputer; with 4 layers on 512 nodes the same computation takes 5 hours 27 minutes.
The DM-HEOM code was designed with portability across different hardware architectures in mind, including GPUs and many-core CPUs. 
It builds on modern programming techniques like OpenCL, C++, and MPI-3 (Message Passing Interface). 
For more details see \cite{Noack2018}.

\subsection{HEOM rate equation}

To connect the HEOM method with a rate matrix formalism following \cite{Jesenko2014,Zhang2016}, it is useful to abbreviate the complete HEOM differential equation by the single operator $\mathcal{L}$:  
\begin{equation}
\frac{\rmd \sigma_{u}}{\rmd t}(t)=-\rmi \mathcal{L}\sigma_{u}(t),
\end{equation}
The projection operator formalism, where the operator
\begin{equation}
\mathcal{P} \sigma_{u}(t) = \sum_{i} \sigma_{0}^{ii}(t) \ket{i}\bra{i}=\sum_i \rho_{ii}(t) \ket{i}\bra{i}
\end{equation}
projects into the diagonal part of the reduced density matrix
and $\mathcal{Q}=1-\mathcal{P}$, 
enables us to write a dynamical equation for the population vector
$\rho_{ii}(t)$ in the form of a generalized quantum-master equation \cite{Jesenko2014}
\begin{equation}\label{eq:Kconv}
\frac{\rmd \rho_{ii}} {\rmd t}(t)= \sum_l^N \int_0^t K_{il} (t-t') \rho_{ll}(t')\rmd t',
\end{equation} 
where the kinetic rate kernels for the HEOM formalism \cite{Zhang2016} are given by
\begin{equation}
{K}_{il} (t)=-\bra{i} \mathcal{P}\mathcal{L} \rme^{-\rmi(\mathcal{Q}\mathcal{L}) t}\mathcal{Q}\mathcal{L} X_l(0) \ket{i},
\end{equation}
and $X_l(0)=\{\rho(0)=\ket{l}\bra{l},\sigma_{u>0}=0 \}$ denotes an initial state of the hierarchy $\sigma_{u}(t)$.\\
A further simplification is achieved by using the Markovian approximation \cite{Jesenko2014} to the kernel rate equation and replace the convolution in Eq. (\ref{eq:Kconv}) by 
\begin{equation}
\frac{\rmd \rho_{ii}} {\rmd t}(t)=\sum_l^N \bar{K}_{il} \bm{\rho}(t),
\label{eq:rateHEOM}
\end{equation}
where 
\begin{equation}
\bar{K}=\int_0^\infty K(t) \rmd t
\end{equation}
represents the integrated rate matrix.

\subsection{Rate equations based on eigenstate representation}

HEOM and standard F\"orster theory are formulated in site representation.
It is also possible to construct rate equations connecting populations of energy eigenstates, based on various Redfield theories.
Common choices are the modified Redfield theory \cite{Yang2002a,Yang2003} and combined (generalized) F\"orster-Redfield theory.
The latter one introduces with $M_{\rm cr}$ a cut-off parameter \cite{Yang2003} to seperate strongly and weakly coupled domains.
For the light harvesting complex II a systematic search and comparison of combined F\"orster-Redfield theory with HEOM results identified as best value of $M_{\rm cr}=30$~cm$^{-1}$ \cite{Kreisbeck2014,Novoderezhkin2017}.
A higher cut-off has been used for PS~I \cite{Yang2003}.

\section{Population dynamics}\label{sec:Pop}

\begin{figure}[t]
\begin{center}
\includegraphics[width=0.7\textwidth]{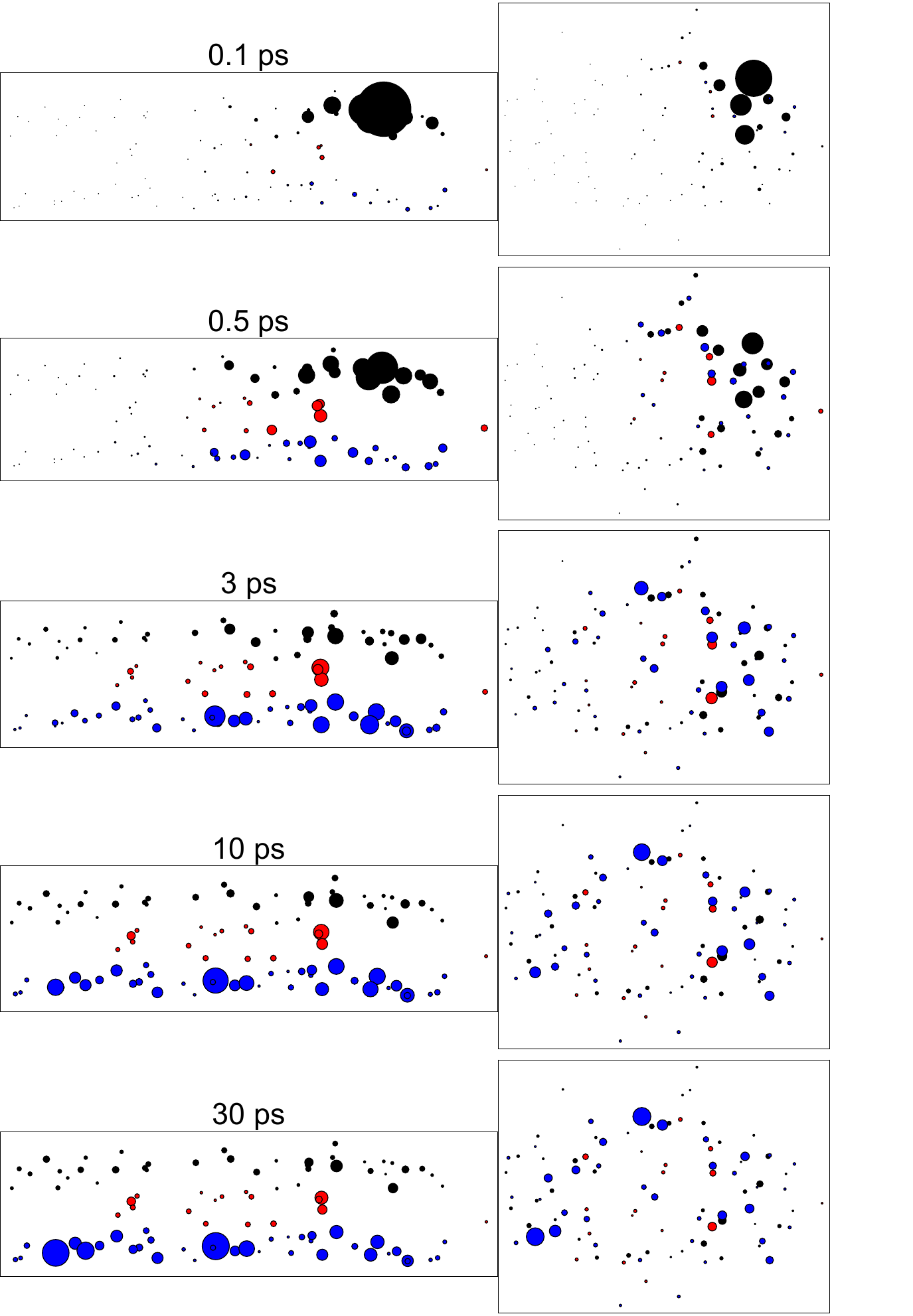}
\end{center}
\caption{Side (left panel) and top (right panel) view of PS~I at different times (orientation and labels given in Fig.~\ref{fig:psi}).
HEOM populations are indicated by the circle area, scaled in proportion to the population. Color code: (blue) lumenal, (black) stromal, (red) middle.
The initial population was placed on pigment PL01.
}\label{fig:heomdots}
\end{figure}

\begin{figure}[t]
\begin{center}
\includegraphics[width=0.7\textwidth]{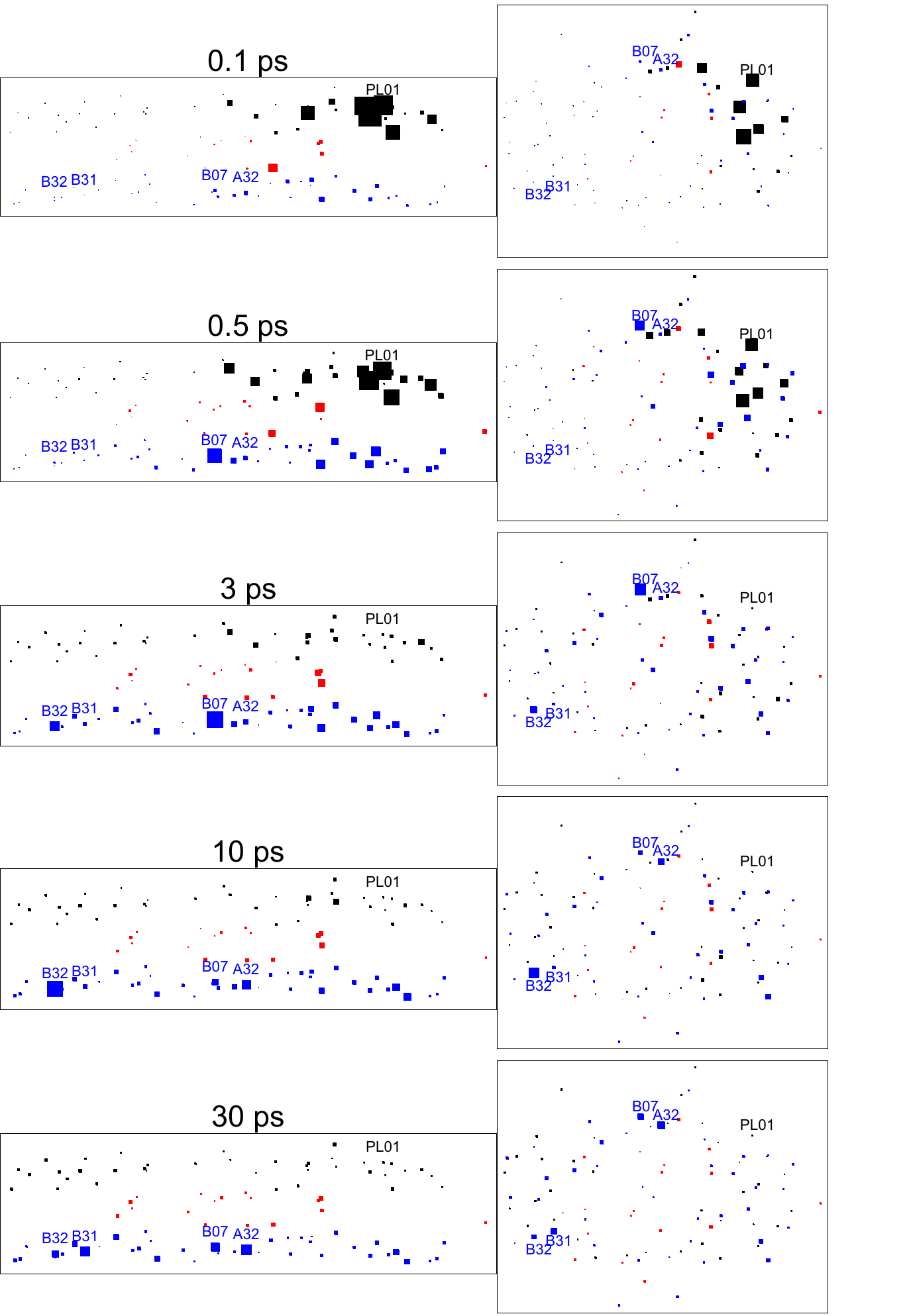}
\end{center}
\caption{Deviations of F\"orster theory from HEOM at different times (Fig.~\ref{fig:heomdots})
The population difference between both methods is shown as the area of the squares.
See also Fig.~\ref{fig:ratevsheomsite} for the populations at selected pigments.
}\label{fig:FoersterErrors}
\end{figure}

To compare HEOM with the different rate equations, we initialize the density matrix at the outer lying PL01 pigment and track the time-evolution of the populations at all pigments (Fig.~\ref{fig:heomdots}). 
A similar initial condition has been used in \cite{Yang2003}. 
We track the energy flow from the outer pigments in the lumenal ring towards the stromal ring and RC of the supercomplex and compare it to other theories.
The difference in energy flow between F\"orster and HEOM theory is shown in Fig.~\ref{fig:FoersterErrors} at different times. 
Main differences appear at time scales around 0.5~ps in the outer region close to the initial pigment and at later times at the red Chlorophyll B~32.

\begin{figure}[t]
\begin{center}
\includegraphics[width=0.49\textwidth]{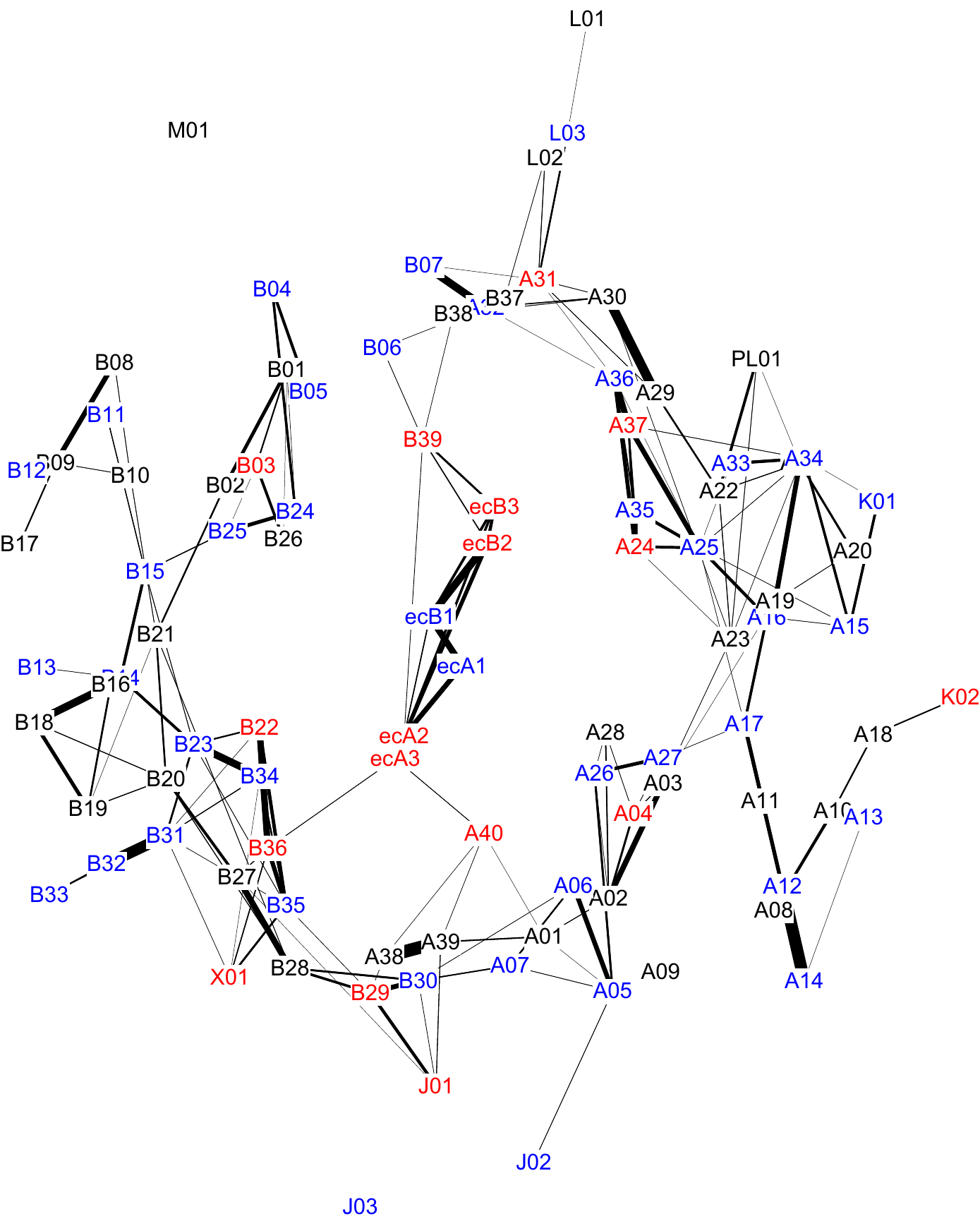}\\
\includegraphics[width=0.49\textwidth]{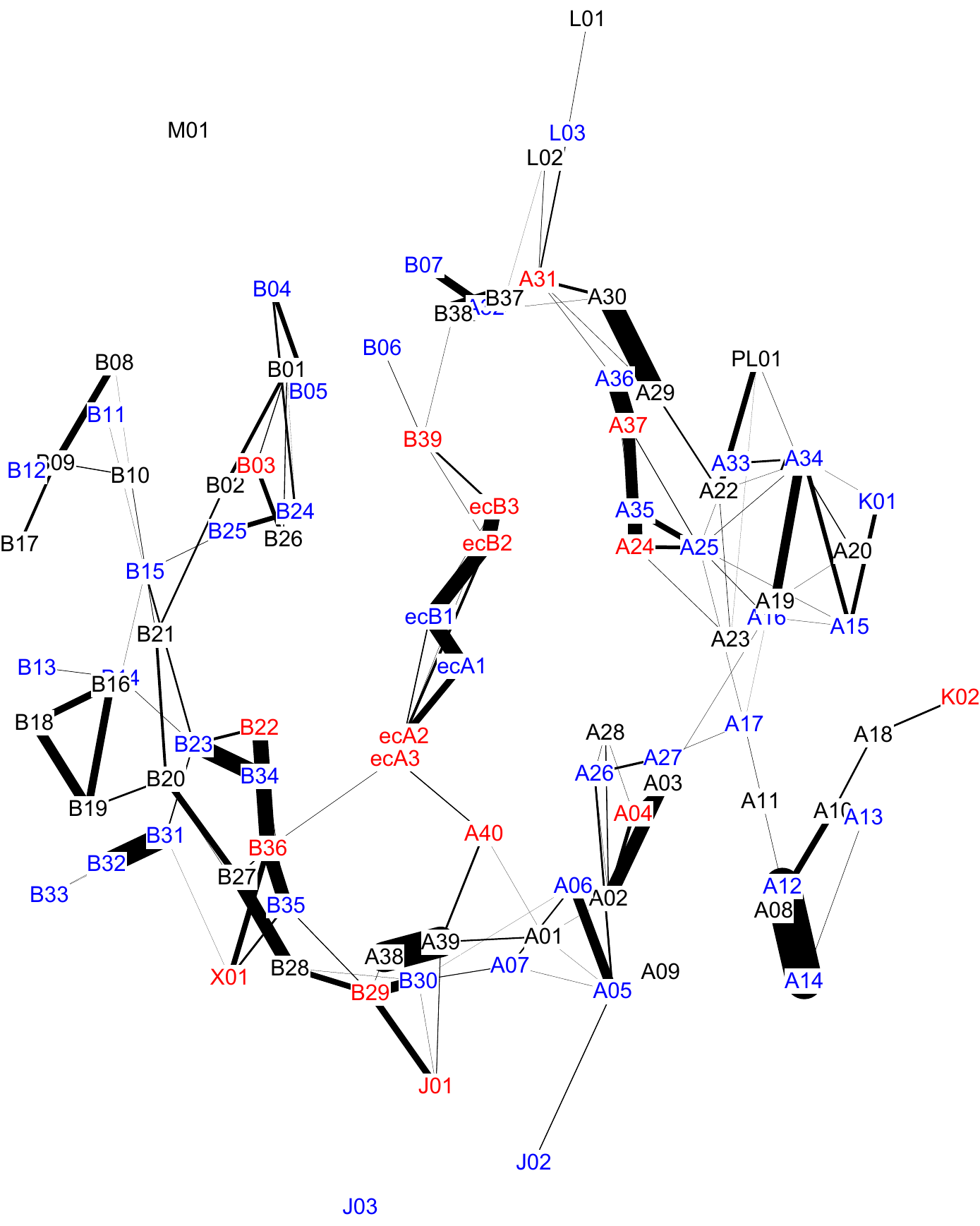}
\end{center}
\caption{(a) Markovian rate HEOM vs (b) F\"orster rates between the PS~I sites. 
The line widths are drawn in proportion to the rate. 
}\label{fig:Network300K}
\end{figure}

Previous calculations \cite{Byrdin2002,Sener2002,Damjanovic2002,Yang2003} had indicated multichannel energy transfer towards the RC.
A scheme of energy transfer pathways within the PS~I complex from {\it Thermosynechococcus elongatus} was first shown in \cite{Byrdin2002}, Fig.~6 therein, based on the protein structure and F\"orster rate equations.
Most transfer rates calculated with F\"oster theory range from $11$~ps$^{-1}$ to $0.3$~ps$^{-1}$ and show the RC well connected to the ring shells via multiple pathways (Fig.~\ref{fig:Network300K}). 
The Markovian approximation to the HEOM rate kernel 
Eq.~(\ref{eq:rateHEOM}) shows a similar multiple pathway transfer, but fails to reach the thermal populations at longer times.

\begin{figure}[t]
\includegraphics[width=0.99\textwidth]{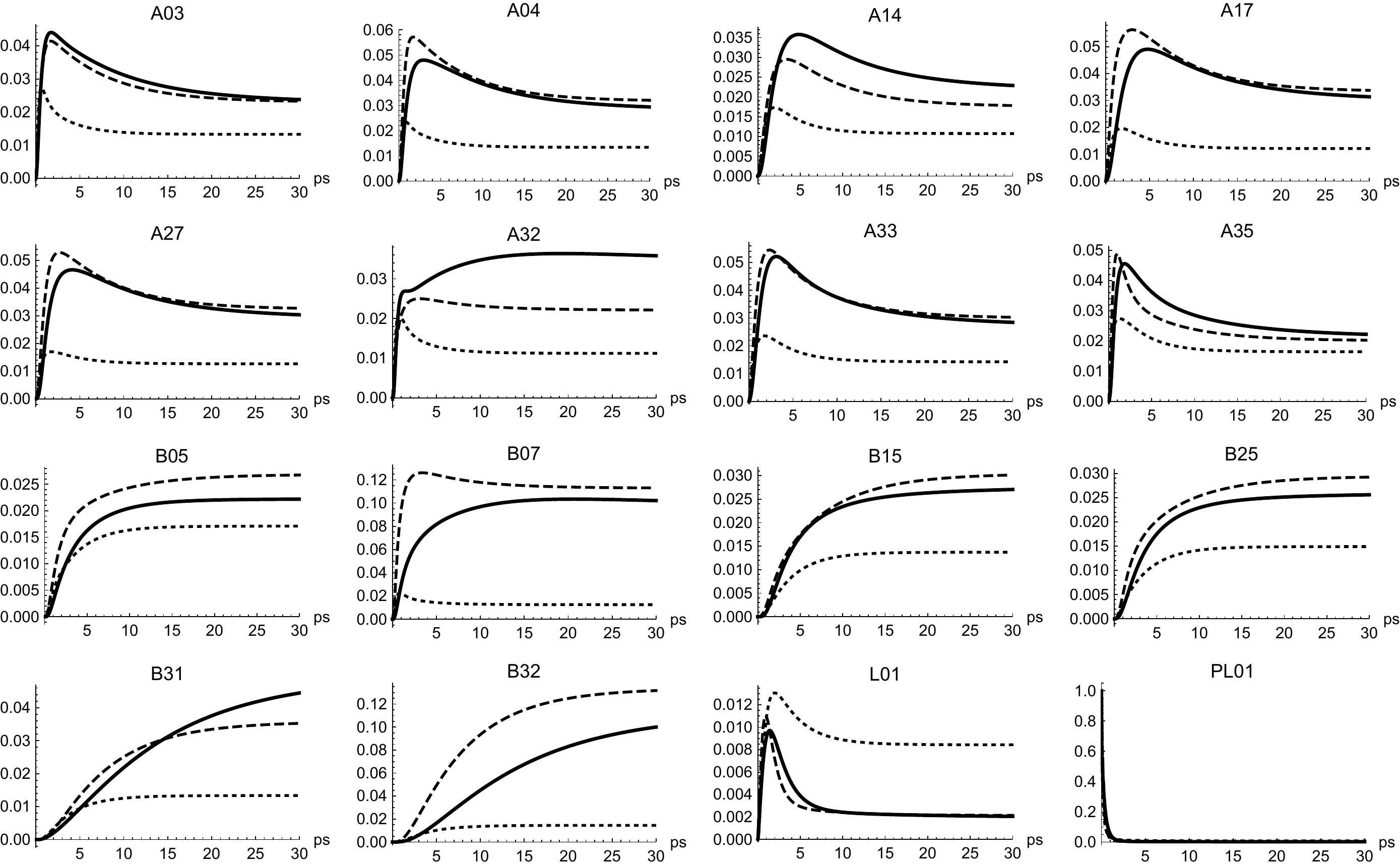}
\caption{Comparison of F\"orster (dashed line), ``Markovian-rate HEOM'' (dotted line), and HEOM dynamics (solid line) at selected pigments.
The initial population was placed on pigment PL01.
}\label{fig:ratevsheomsite}
\end{figure}

Fig.~\ref{fig:ratevsheomsite} compares the populations at selected pigments obtained with HEOM versus F\"orster theory, and the Markovian rate HEOM.
In particular the Markovian rate solution differs considerably from the exact HEOM solution.
For the Fenna Matthews Olson complex (not shown here), the comparison of the Markovian rates derived from HEOM with the exact dynamics are in better agreement.
We attribute this different level of agreement to the different regimes of energy transfers in both complexes.
The Fenna Matthews Olson complex operates closer to the secular Redfield regime \cite{Kramer2018a}, while the PS~I complex is more amendable to F\"orster theory.

\begin{figure}[t]
\begin{center}
\includegraphics[width=0.49\textwidth]{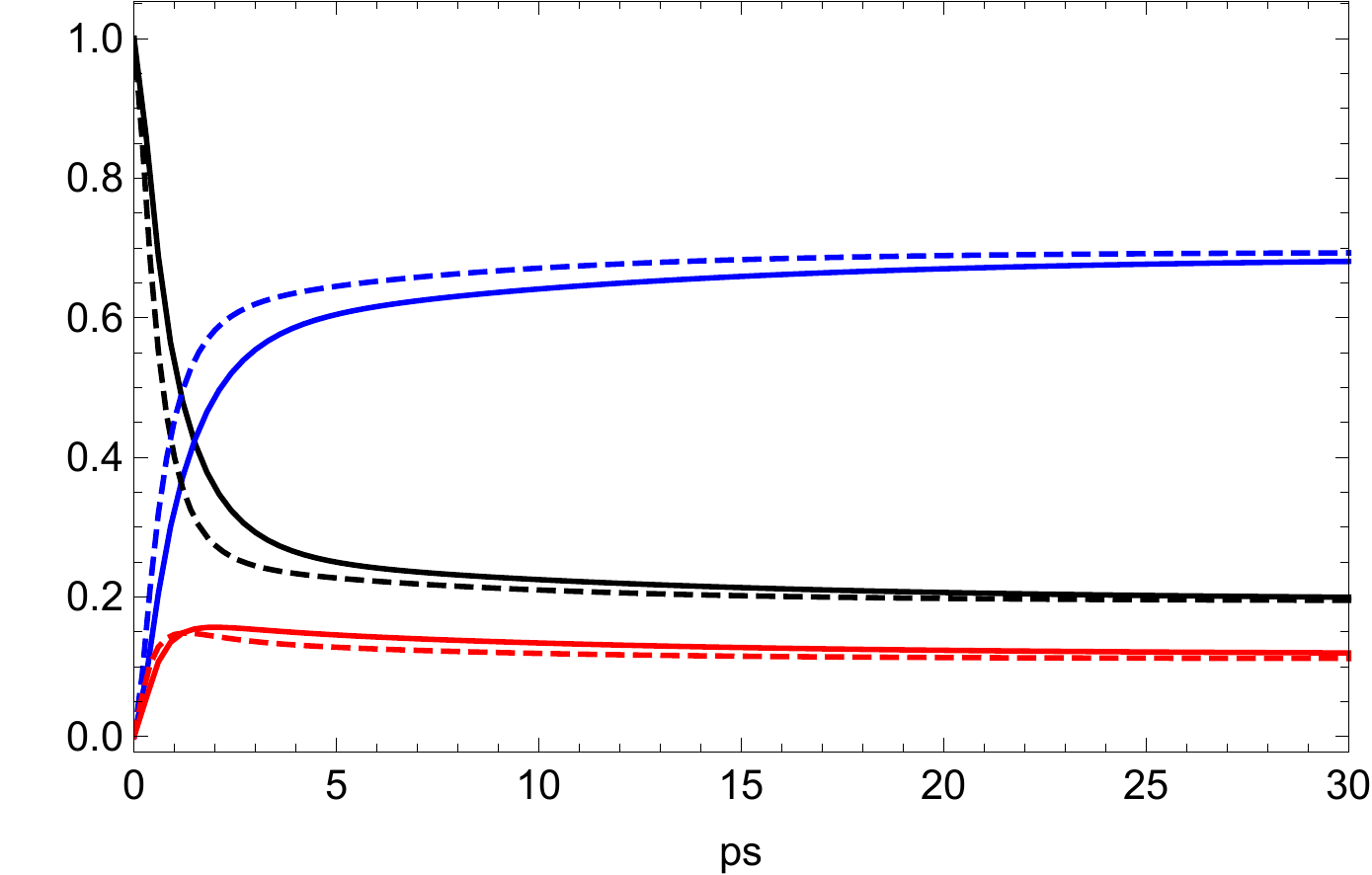}
\includegraphics[width=0.49\textwidth]{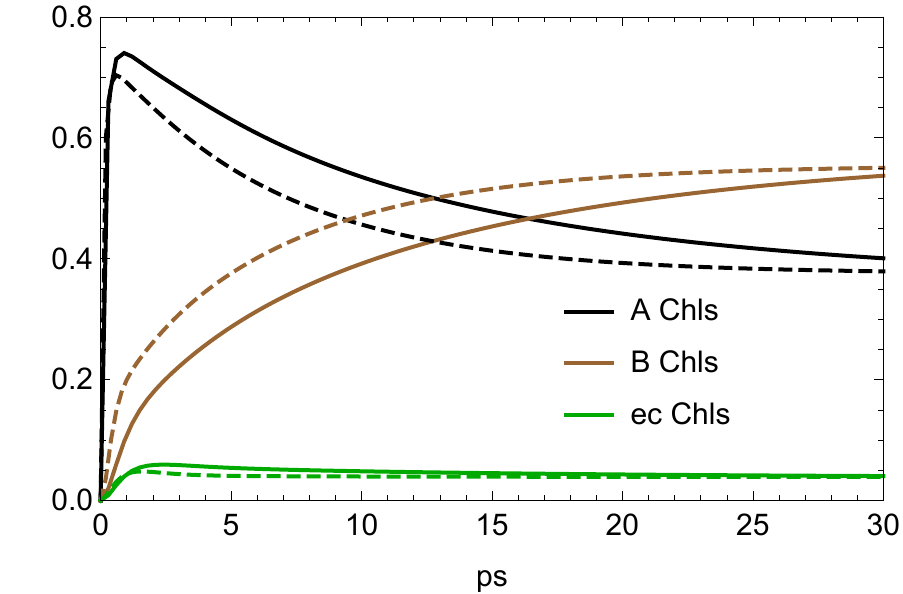}
\end{center}
\caption{Comparison of F\"orster (dashed lines) and HEOM (solid lines) populations aggregated across (a) (black) stromal / (blue) lumenal / (red) middle pigments, 
(b) the A/B branches.
The HEOM calculation shows a slower transition of A to B chlorophylls compared to the F\"orster rate expectation.
The initial population was placed on PL01.
}\label{fig:ratevsheomSML}
\end{figure}

An interesting open question in PS~I is the role of the A/B branches in the electron transfer dynamics \cite{Ramesh2004,Badshah2017} towards the RC.  
The cross-over between the A/B branch occurs in the DM-HEOM calculation at $17$~ps, while the F\"orster rate puts it at $9$~ps (Fig.~\ref{fig:ratevsheomSML}).
This indicates that the exciton transfer in PS~I proceeds slower than predicted by F\"orster theory.

The difference between F\"orster and HEOM theory is shown in Fig.~\ref{fig:FoersterErrors} at different times and shows that in particular the lowest-energy chlorophyll B32 is populated at later time according to the HEOM theory compared to the F\"orster prediction (see also the B32 panel in Fig.~\ref{fig:ratevsheomsite}). 
Also the B07 and A32 chlorophylls differ at intermediate times in  population between both theories, with B07 being another low energy Chl.

\begin{figure}[t]
\includegraphics[width=0.99\textwidth]{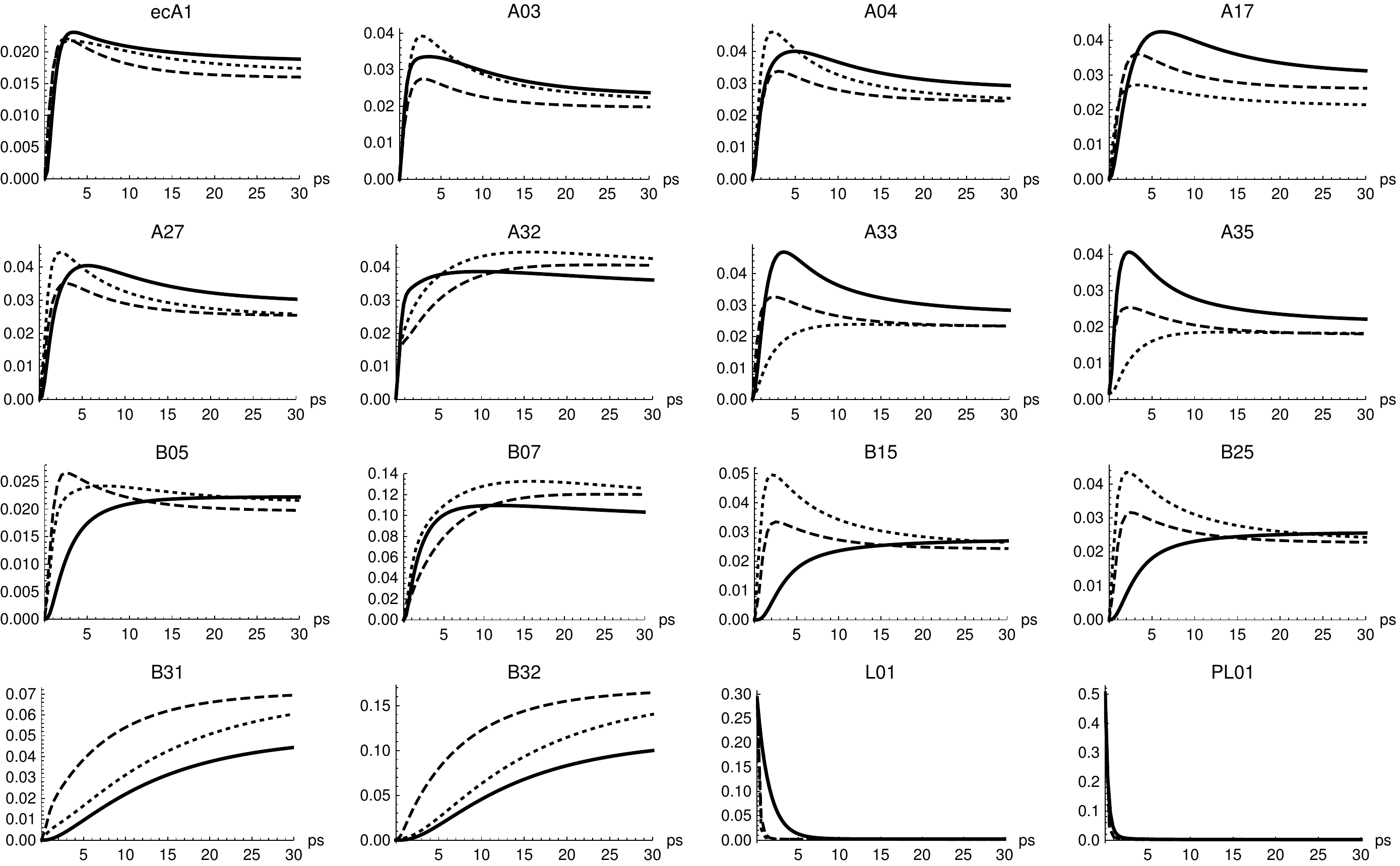}
\caption{Comparison of modified Redfield (dashed line), combined F\"orster-Redfield (dotted line, $M_{\rm cr}=70$~cm$^{-1}$), and HEOM dynamics (solid line) at selected pigments.
The initial population corresponds to the eigenstate with the largest population on pigment PL01.
}\label{fig:ratevsheomsiteMF}
\end{figure}

Finally, we contrast HEOM results with modified Redfield and combined F\"orster-Redfield theories.
To facilitate the comparison, we initialize the populations dynamics as a single energy eigenstate of the diagonalized exciton Hamiltonian 
\begin{equation}\label{eq:ee}
H^{EE}=A H_{\rm ex,site} A^T, H^{EE}={\rm diag}(E_1,\ldots E_{96})
\end{equation}
The eigenstate corresponding to $E_{69}$ (energies sorted from lowest to highest value) has the largest overlap (0.503) with the PL01 pigment.
HEOM is initialized with the density matrix in site basis corresponding to
$|E_{69}\rangle\langle E_{69}|$.
The eigenstate populations of the other methods are transformed to site populations by the inverse relation of Eq.~(\ref{eq:ee}).
Fig.~\ref{fig:ratevsheomsiteMF} shows no improvement of the description of the dynamics by using these approaches compared to the level of standard F\"orster theory.

\section{Spectra}
\label{sec:Spectra}

\subsection{Linear Absorption}

\begin{figure}[t]
\begin{center}
\includegraphics[width=0.5\textwidth]{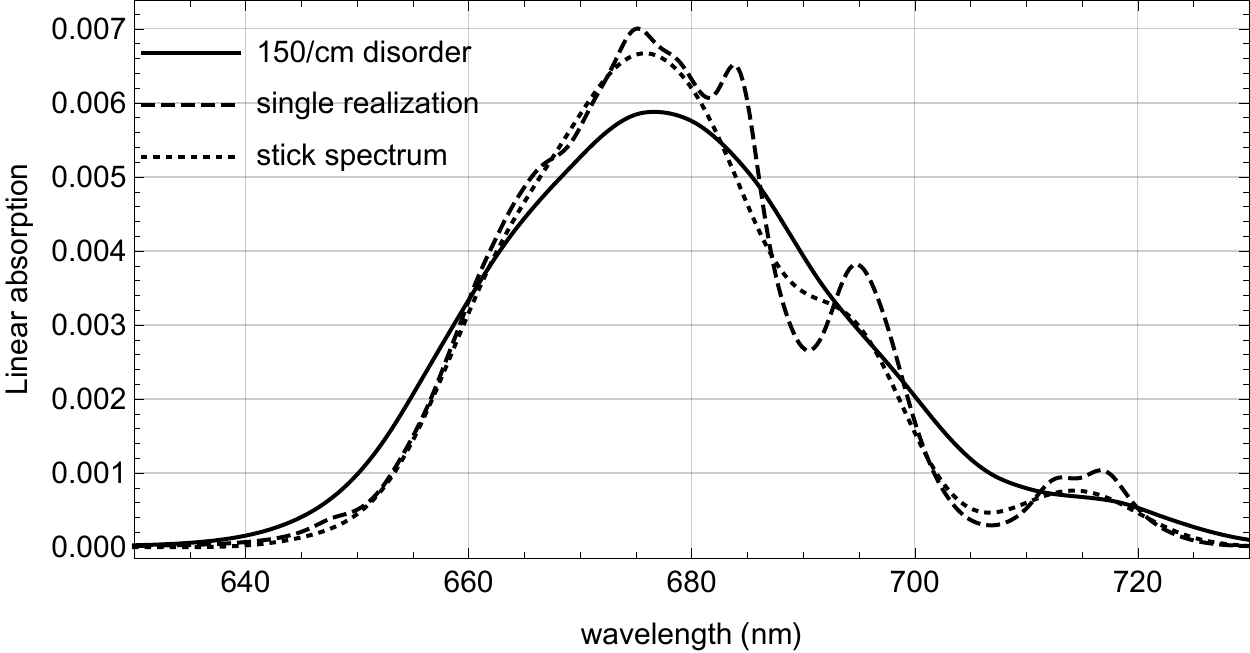}
\end{center}
\caption{Linear absorption spectra of PS~I at $T=300$~K.
Disorder averaged (1000 realizations, $\sigma=150$~cm$^{-1}$), (dashed) single realization linear absorption spectra at $T=300$~K, (dotted) Gaussian broadened stick spectrum.
The energies are shifted by 2300~cm$^{-1}$.
}\label{fig:LA300K}
\end{figure}

The linear absorption spectra in Fig.~\ref{fig:LA300K} are calculated with DM-HEOM following \cite{Kramer2018a}, Eq.~(56).
The transition dipoles of the $Q_y$-band are assumed to be oriented along the nitrogen $N_B$-$N_D$ positions in the optimized 1JB0 structure \cite{Yin2007}, while we ignore the $Q_x$ band.
To assess the impact of static disorder we consider both, a single realization with the Hamiltonian \cite{Yin2007}, and in addition the ensemble average of 1000 calculations with uncorrelated diagonal disorder added to the site energies (standard deviation $150$~cm$^{-1}$), to match the experimental spectra at $T=295$~K.
The TD-DFT site energies are shifted by 2300~cm$^{-1}$ to the location of the experimental absorption spectra \cite{Byrdin2002}.
The room temperature PS~I absorption spectrum consists of a broad main antenna ($700$-$645$~nm) absorption band and a red shoulder below $715$~nm.
The lowest-energy band of the reaction center is at $\approx 700$~nm.
The `red' absorption band which extends below the RC absorption is a unique feature of the PS~I complex and varies among the cyanobacterial and plant PS~I complex \cite{Gobets2001}, Fig.~2.
For comparison, we also include the Gaussian broadened stick spectrum (\cite{Yin2007}, Fig.~1)
obtained at 0~K ignoring all intermolecular and intramolecular relaxation.  
This extremely computationally efficient approach yields a qualitatively realistic description of the absorption spectrum but provides no information concerning the dynamics of energy transfer or any other possible photochemical process.
Basically, this result echoes the known feature that inhomogeneously absorption spectra in themselves often reveal little information related to dynamical processes; however, it is necessary that any method such as DM-HEOM that is designed to model dynamics realistically also adequately describes absorption.

\subsection{Circular and Linear Dichroism}

\begin{figure}[t]
\begin{center}
\includegraphics[width=0.5\textwidth]{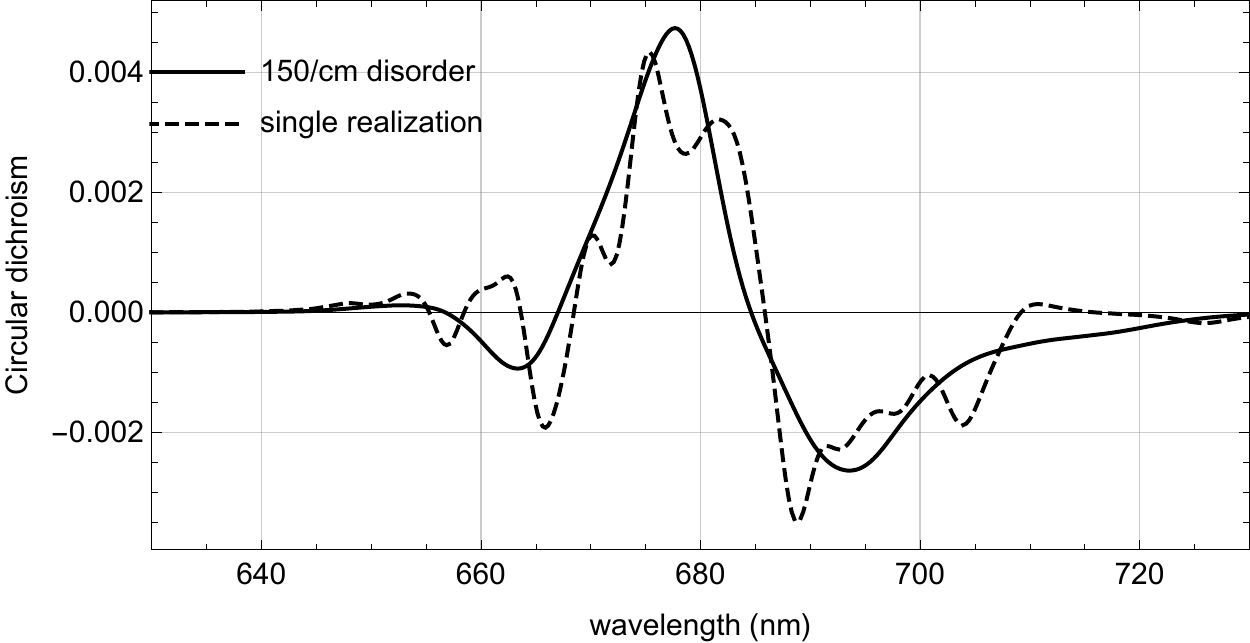}
\end{center}
\caption{Circular dichroism of PS~I at $T=300$~K.
Disorder averaged (1000 realizations, $\sigma=150$~cm$^{-1}$), (dashed) single realization circular dichroism at $T=300$~K.
The energies are shifted by 2300~cm$^{-1}$.
}\label{fig:CD300K}
\end{figure}

The circular dichroism (CD) spectra are shown in Fig.~\ref{fig:CD300K}, calculated with DM-HEOM at $T=300$t~K using \cite{Kramer2018a}, Eq.~(59).
We do not consider the non conservative part of the PS~I circular dichroism, which is visible in the experimental spectra (Fig.~8C, \cite{Byrdin2002}).
The simulations show a negative dip at 660~nm, which is absent in the experiments, but seen  in other theoretical simulations \cite{Byrdin2002,Abramavicius2009a}.

Other features of the experimental CD spectra are well reproduced, including the knee around $700$~nm.
Disorder has a large impact on the CD spectra as seen by comparing the ensemble averaged result with the single realization case.

\begin{figure}[t]
\begin{center}
\includegraphics[width=0.5\textwidth]{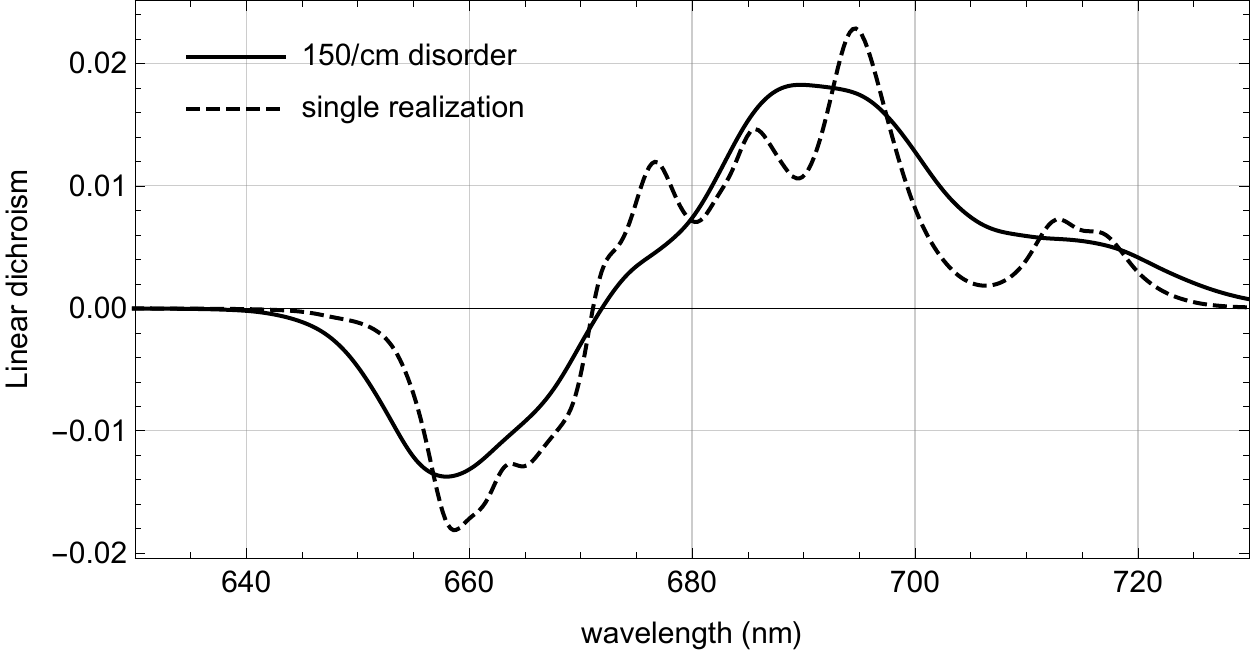}
\end{center}
\caption{Linear dichroism of PS~I at $T=300$~K.
Disorder averaged (1000 realizations, $\sigma=150$~cm$^{-1}$), (dashed) single realization linear dichroism at $T=300$~K.
The energies are shifted by 2300~cm$^{-1}$.
}\label{fig:LD300K}
\end{figure}

In addition we evaluated the linear dichroism (Fig.~\ref{fig:LD300K}) using Ref.~\cite{Prokhorenko2003}, Eq.~9.3, under the assumption of a perfectly aligned structure in the $x-y$ plane and the molecular symmetry axis taken along the $z$-axis.
The minimum is found around 660~nm in accordance with the experimental data  \cite{Byrdin2002}, Fig.~3A.

\section{Conclusion}\label{sec:Discussion}

We have used the DM-HEOM to compute the energy flow in the PSI system and compare it with simplified approaches.
The site energies and excitonic couplings are taken from the DFT results and no fitting to experimental spectra has been performed.
The analysis of exciton flow in PS~I with DM-HEOM reveals a slow transfer, even compared to F\"orster rates.
The HEOM computation validates previous simplified approaches relying on F\"orster rates, with a priori unknown error bounds \cite{Byrdin2002}.
A simplified rate model based on a Markovian approximation of HEOM does not yield good agreement with the exact calculation for PS~I.
The applicability of the Markovian rate HEOM and F\"orster theory warrants further theoretical investigation.
For PS~I we find that the combined F\"orster-modified Redfield approach does not improve the dynamics with respect to the standard F\"orster treatment.
The experimental linear absorption and circular dichroism spectra of PS~I are partially reproduced by DM-HEOM at $T=300$~K, with the exception of a negative dip at higher energies in the CD spectra. 
However, static spectra are not directly reflecting the dynamics and energy transfer.
In future work, we plan to use DM-HEOM to compute time-resolved spectra to track the energy flow throughout the complex.

\section*{Acknowledgements}

This contribution is dedicated to Prof.~Wolfgang Domcke whose work on time-dependent quantum mechanics and time-resolved spectroscopy provides the required tools and methods to analyze chemical reaction dynamics.

We thank G.~Laubender, Y.~Zelinskyy, and Th.~Steinke for helpful discussions.
The work was supported by the German Research Foundation (DFG) grants KR~2889 and RE~1389 (``Realistic Simulations of Photoactive Systems on HPC Clusters with Many-Core Processors'') and the Intel Research Center for Many-core High-Performance Computing at ZIB.
We acknowledge compute time allocation by the North-German Supercomputing Alliance (HLRN).
M.R.\ has received funding from the European Union's Horizon 2020 research and innovation programme under the Marie Sklodowska-Curie grant agreement No.~707636.

\clearpage

\end{document}